\documentclass[twocolumn,preprintnumbers,amsmath,amssymb,superscriptaddress]{revtex4}

\usepackage{graphicx}
\usepackage{dcolumn}
\usepackage{bm}
\usepackage{color}
\def\mcl{}

\def\rice{Department of Physics and Astronomy, MS 108, Rice University, Houston, Texas 77005, USA}
\def\llnl{Lawrence Livermore National Laboratory, Livermore, California 94551, USA}
\def\dd{$^\circ$}

\def\pt{peak}
\def\tI{$ \theta_{17}\ $}
\def\tII{$ \theta_{30}\ $}
\def\tIII{$ \theta_{45}\ $}
\def\iI{$ a_1\ $}
\def\iII{$ a_{3} \ $}
\def\iIII{$ a_{10}\ $}
\def\cI{ 300 fs }
\def\cIII{ 1.4 ps }
\def\ecI{$ 0.1 \ n_{cr} \ $}
\def\ecII{$ n_{cr} \ $}
\def\per{$ P_e/I_0 \ $} 
\def\wal{Wilks \textit{et al.}}

\begin{document}

\title{Focusing of Intense Subpicosecond Laser Pulses in Wedge Targets}

\author{M. C. Levy}
\affiliation{\rice}
\affiliation{\llnl}
\author{A. J. Kemp}
\author{S. C. Wilks}
\author{L. Divol}
\affiliation{\llnl}
\author{M. G. Baring}
\affiliation{\rice}

\date{\today}

\begin{abstract}
Two dimensional particle-in-cell simulations characterizing the interaction of ultraintense short pulse lasers in the range $10^{18} \le I \le 10^{20}\ W/cm^2$ with converging target geometries are presented. Seeking to examine intensity amplification in high-power laser systems, where focal spots are typically non-diffraction limited, we describe key dynamical features as the injected laser intensity and convergence angle of the target are systematically varied. We find that laser pulses are focused down to a wavelength with the peak intensity amplified by an order of magnitude beyond its vacuum value, \mcl{and develop a simple model for how the peak location moves back towards the injection plane over time}. This performance is sustained over hundreds of femtoseconds and scales to laser intensities beyond $10^{20} \ W/cm^2$ at $1 \mu m$ wavelength.
\end{abstract}

\maketitle

\section{Introduction} 

In recent years advances in short pulse laser technology and focusing techniques such as chirped-pulse amplification (CPA)\cite{Strickland1985} have enabled production of petawatt laser systems with intensities exceeding $10^{20}\ W/cm^2$, and pulse lengths on the order of hundreds of femtoseconds.  The fields generated by such ultraintense pulses cause electrons to oscillate rapidly, with relativistic effects becoming appreciable for $p_{osc}/m_ec \ge 1$, where $p_{osc}$ is the transverse relativistic electron quiver momentum in the laser  fields.  The dimensionless laser parameter $a \equiv e E_0/(m_e \omega_0 c) = p_{osc}/m_ec \ge 1$ defines this bound, where $E_0$ and $\omega_0$ are the peak laser electric field and frequency respectively. Pulses in this regime enable compelling laboratory experiments in high-energy density physics (HEDP), with applications to diverse fields such as antimatter creation\cite{Chen2009b,Fedotov2010a}, high-energy astrophysics\cite{Remington2006}, and inertial confinement fusion (ICF.)\cite{Tabak1994}

It is well-established that the interaction of lasers at ultrahigh intensities $I \lambda^2 \ge 1.37 \times 10^{18} W  \mu m^2/cm^2$ with solid targets results in the production of energetic electrons\cite{Wilks1992}.  The impinging laser propagates into the target until it reaches an interface with density $n\approx \gamma\times n_{cr}$, where $n_{cr}=m_e \omega_{0}^2/(4\pi e^2)$, the critical density at which the laser frequency equals the electron plasma frequency; and $\gamma = \sqrt{1+a^2}$ is the relativistic factor corresponding to electron oscillation in the transverse laser fields. For idealized planar targets irradiated at normal incidence,Êthe system dynamics can be characterized in a general sense by two properties of the laser: the light pressure $P\approx 2I/c$, and relativistic ponderomotive force $ \vec{f_p} = \vec{\nabla} (\gamma_t-1) m_e c^2$.  Here $\gamma_t=\sqrt{1+(p_e/m_e c^2)^2}$ corresponds to the relativistic electron quiver momentum in both the transverse and longitudinal fields. The light pressure, which can exceed Gbar for $I\sim 10^{19}W/cm^2$, pushes on ions at the critical interface and effectively bores a hole into the target. The time-averaged component of $\vec{f_p}$ acts to further steepen the critical interface, while the oscillatory component sets up an effective electrostatic field, heating electrons and nonadiabatically injecting them into the solid target at frequency $2\omega_0$\cite{Kruer1984}. The ponderomotive scaling\cite{Wilks1992} of the hot electron temperature, given by $\epsilon_h=k\ T_h=m_e c^2 ( \gamma -1)$\label{eqn:pondT}, represents a widely-adopted feature for ultraintense laser-plasma interaction (LPI) research.

Short pulse laser applications such as $K_{\alpha}$ based X-ray backlighters require relativistic laser intensities $>10^{19}\ W/cm^2$ over relatively small spots, i.e., radius $\sim10 \mu m$, while high power laser systems will deliver most of their energy in $\sim200 \mu m$ spots.  Much experimental and computational research over the past several years has explored the potential utility of convergent geometry targets to amplify laser intensity, as well as to improve particle guiding, temperature and distribution\cite{Ping2008,Cottrill2010,Zeng2010,Yu2010}. In the fast ignitor (FI) approach to ICF, stringent requirements on target symmetry are relaxed and  significantly higher gain can be achieved by decoupling the processes of fuel compression and ignition, with an ultraintense picosecond laser pulse most commonly providing the ``spark'' \cite{Tabak1994}.  In one ignition scheme, a re-entrant cone is embedded into the precompressed fuel to shield the ignitor pulse from coronal plasma created during fuel compression, and to allow immediate access to the dense core.  There is currently much research being done to understand the complex, multi-scale physical processes at work in this system, with outstanding issues in areas such as laser transmission and focusing in the cone, laser energy coupling into hot electrons, and electron divergence in the dense thermonuclear fuel\cite{Kemp2010a}. Furthermore, in the context of proton acceleration there is currently much active research in novel target geometries, such as funnel and flat-top cones. Recent experiments on the Trident laser at Los Alamos National Laboratory have shown these targets to increase laser-particle energy coupling by nearly 5 times and to enhance laser focusing towards the cone tip\cite{Flippo2008}.   

In this paper, we seek to characterize the focusing and evolution of ultraintense short pulse lasers in converging target geometries. To distill and examine key physical processes, we use particle-in-cell (PIC) simulations to model wedge targets in the limit of no prepulse, i.e. high-contrast laser pulse propagation in a cleared channel.  The effects of prepulse in a closed cone, for a single angle, were reported on in MacPhee \textit{et al.}\cite{MacPhee2010a}  Recent numerical work published by other groups has indicated that peak laser focusing of $\sim$15 times could be achieved in a small spot size of radius $\sim 1 \lambda_0$, using hollow tip cone (wedge) targets.\cite{Zeng2010} It was proposed that the intense, highly-focused beams generated by these configurations could have applications for HEDP fields such as $K_\alpha$ based backlighting. However, these simulations were performed in the limit of an optical physics model with static boundaries\cite{Zeng2010}, and further research has considered evolving weakly overdense target walls with lasers of intensity $I<10^{20} \ W/cm^2$ over $\sim$350 fs \cite{Yu2010}. The results presented in this paper take into account the additional effects of hot electron generation and phase space evolution, and describe nonlinear laser-plasma interaction in $100\times n_{cr}$ targets over picosecond timescales. We consider a pulse-target parameter space that represents a sampling of recent literature\cite{Kemp2010a,MacPhee2010a,Zeng2010}, and is comprised by $1 \mu m$ wavelength lasers with intensities in the range  $10^{18}-10^{20} \ W/cm^2$ and targets with half-angles of 17\dd, 30\dd and 45\dd. We identify key dynamical features and trends as laser intensity and target angle are systematically varied, such as the generation of a dominant hot electron filament aligned with the target tip for $a>1$ pulses in the 17\dd \ target. In these configurations particle acceleration mechanisms in the underdense plasma, most notably self-modulated laser wakefield acceleration\cite{Esarey2009,Pukhov2003a} (SM LWFA), excite electrons to several times the ponderomotive potential. By contrast the wide 45\dd \ target exhibits two equally dominant electron filaments off to the sides of the tip, with electron acceleration through the tip effectively suppressed. We interpret this trend in the context of the relative efficiencies of ponderomotive self-channeling\cite{Pukhov2003a} and collisionless laser energy absorption mechanisms such as Brunel heating\cite{Brunel1987} near the critical interface.  For pulses with intensity higher than $I_{19}$ in the 45\dd \ target, over time the laser pressure becomes insufficient to clear plasma ablated from the target walls from the laser channel. This results in effective closure of the target tip. 

The 17\dd $ - I_{20}$ pulse-target configuration is identified as optimal with peak laser focusing $I_{\pt}/I_0\approx 8.8$ and maximal electron flux density \per $\approx 3.2$ at the target tip. This $I_{\pt}$ represents an 82\% focusing efficiency compared to an ideal optical simulation with static boundaries, with losses primarily going into electron heating. Also observed in this geometry is a clear trend where the region of peak laser intensity regresses away from the target tip at intensity-dependent rates that saturate at $I\gg 10^{19} \ W / cm^2$. Interestingly, this trend suggests the 17\dd \ geometry may be representative of a broader class of pulse-target configurations that are characterized by a sustained dynamical equilibrium between plasma expansion and laser radiation pressure. A general analytical model is developed for these systems that predicts the front velocity $u_f$ and yields key scalings $I_{\pt}, P_e \propto a^2$, in good agreement with the 17\dd \ target results.  This unique class of wedge target configurations has highly desirable properties and may have applications for the design of $K_\alpha$ based X-ray backlighters to image ICF implosions and other HEDP experiments.\cite{Park2006}. In this context they represent experimentally simple systems to achieve bursts of 10x boosted light for hundreds of femtoseconds, focused down to small spots of $\sim 1 \lambda_0$.  As X-ray image resolution is proportional to source size, this small spot may enable high resolution radiography, while the amplified laser intensity drives a higher energy flux over a fixed time interval for the snapshot. Because the $I_{\pt}$ and $P_e$ scalings extend to the regime $I \gg I_{20}$, these configurations may become even more attractive as higher intensity short pulse laser systems such as NIF-ARC\cite{Key2007} come online in the future.

\section{Numerical Modeling of Laser Focusing in Wedge Targets}

We use the particle-in-cell code PSC\cite{RuhlPsc}, which employs the Finite-Difference-Time-Domain (FDTD) numerical scheme to move particles according to the Lorentz force equation and evolve fields according to Maxwell's equations on a discrete grid.  We present the results of nine simulations, representing all combinations of pulse intensities $I_0=1.37 \times \{ 10^{18}, 10^{19}, 10^{20} \} \ W/cm^2$ corresponding to $a = \{1, 3.16, 10 \}$ at wavelength $\lambda_0=1\mu m$, and target opening half-angles of 17, 30, and 45\dd.  For convenience we define the pulse-target notation $a_n  \theta_m$ where $n, m$ index the dimensionless laser parameter and target-half angle in degrees. Selection of the \tI target is motivated by recent optical simulations that indicate peak laser focusing could reach 15 times in 16-18\dd \ hollow tip cone (wedge) targets\cite{Zeng2010}. Note that irrespective of angle the targets have a fixed height, thickness, and tip width of $30\lambda_0$, $4.2\lambda_0$, and $\Delta y_{tip}=5\lambda_0$ respectively. Particles are represented with 120 electrons and 30 ions per cell, yielding a total of $\sim 10^7$ quasi-particles initially. Ions are efficiently modeled using fewer particles due to their high inertia and particle weighting is adjusted to preserve charge neutrality. Radiating and reflecting boundary conditions are used for particles and fields respectively, where particles incident upon the boundary have the component of their momenta normal to the boundary inverted. Our simulations are two dimensional Cartesian with system size $50\lambda_0 \ \times$ 40--84$\lambda_0$ in the longitudinal--$\hat{z}$ and transverse--$\hat{y}$ directions respectively, comprising a mesh of 1600 $\times$ 1280--2688 cells with uniform grid size $\Delta y=\Delta z=0.031\lambda_0$. The simulation duration is $550 \tau_0$, with $\tau_0 =\lambda_0/c \approx 3.3$ fs and timestep $\Delta t \approx 0.045$ fs as determined by the Courant criterion\cite{Birdsall1991}.

\begin{figure}[h]
\begin{center}
\resizebox{8.2cm}{!}{\includegraphics{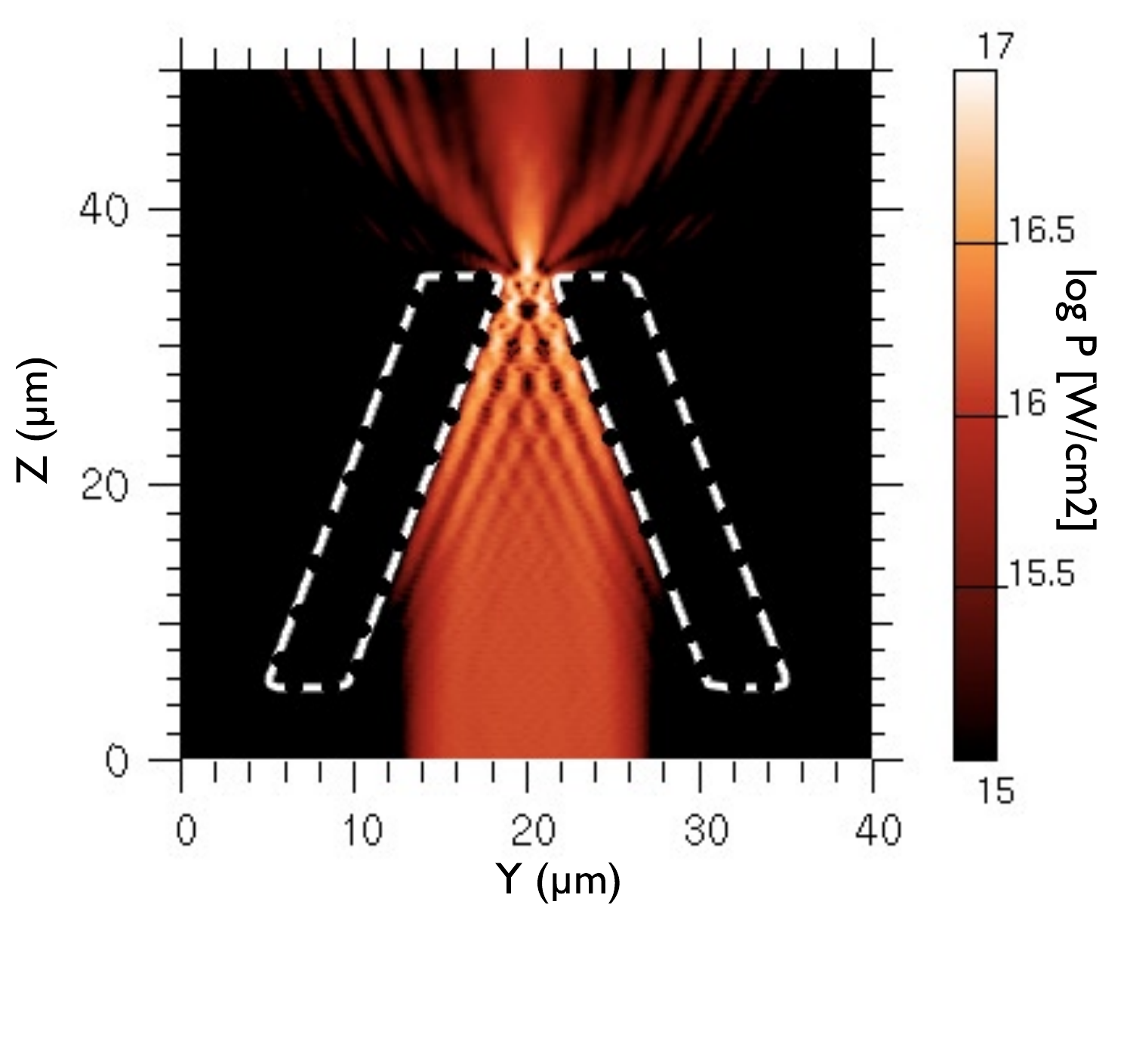}}
  \end{center}
  \caption{Schematic for the \tI target configuration with the low intensity $a=0.1$ reference beam (see text); laser longitudinal Poynting flux $P_z$ shows $I_{\pt}/I_0$=10.8 peak focusing with superimposed white dashed lines indicating initial target position.}
\label{fig:setup} 
\end{figure}

Laser pulses in the simulations presented here are super-Gaussian in the transverse direction with $I(y)=I_0\times \exp \ -(y/y_0)^8$. Temporal profiles are Gaussian and rise over 10 optical cycles with a semi-infinite pulse envelope. Fig. \ref{fig:setup} shows the laser propagating upward into the system from the injection plane at $z=0$, with the target tip at $z_{tip}=35 \lambda_0$. The laser electric field is linearly polarized in the simulation plane. The beam half-width half-maxima (HWHM) are $w_0=\{6.0, 9.9, 15.1\}\times\lambda_0$ for targets half-angles  \tI, \tII and \tIII respectively, a constant fraction of the target base width. Note that we take $w(z) = w_0$, as the expected beam divergence is negligible over the target longitudinal height. Consider an analogous Gaussian beam with waist $w_0$, where we see the vacuum diffraction length $z_r=\pi w_0^2/\lambda_0 \gg z_{tip}$.  

Wedge targets are modeled as two converging $100\times n_{cr}$ plasma slabs with a step-function density profile, initial temperature $T_e,T_i=1$ keV, ions of  $M_i=3672\ m_e$, and charge state $Z=1$. For subpicosecond timescales the $Z/A=1/2$ ratio here accurately reflects the ionization state of high-Z composition targets near the critical interface. In the region of laser irradiation system dynamics are dominated by collisionless processes, as $\nu_{ei}/\omega_0 \sim 6\times10^{-2}$ for the Spitzer frequency $\nu_{ei} = 8\pi n_e e^4 \ln \Lambda/[3 \sqrt{3 m_e} (k T)^{3/2}]$\cite{Ginzburg1970}. Resistive processes in the solid density target walls are neglected, as our analysis is primarily concerned with characterizing laser focusing and plasma ablation near the critical interface. The classical plasma skin depth at oblique incidence $l_s\approx c/\omega_p \cos \theta_T$ is resolved $\Delta z/l_s \sim 1$ for all targets, where the angle of incidence $\theta_T$ is the complement of the target opening half-angle. We have performed a separate test simulation with two times higher grid resolution in the \iII \tIII configuration and found no major qualitative differences. 

\section{Characterizations of Dynamical Pulse-Target Evolution}

\begin{figure}[h]
\begin{center}
\resizebox{8.2cm}{6.5cm}{\includegraphics{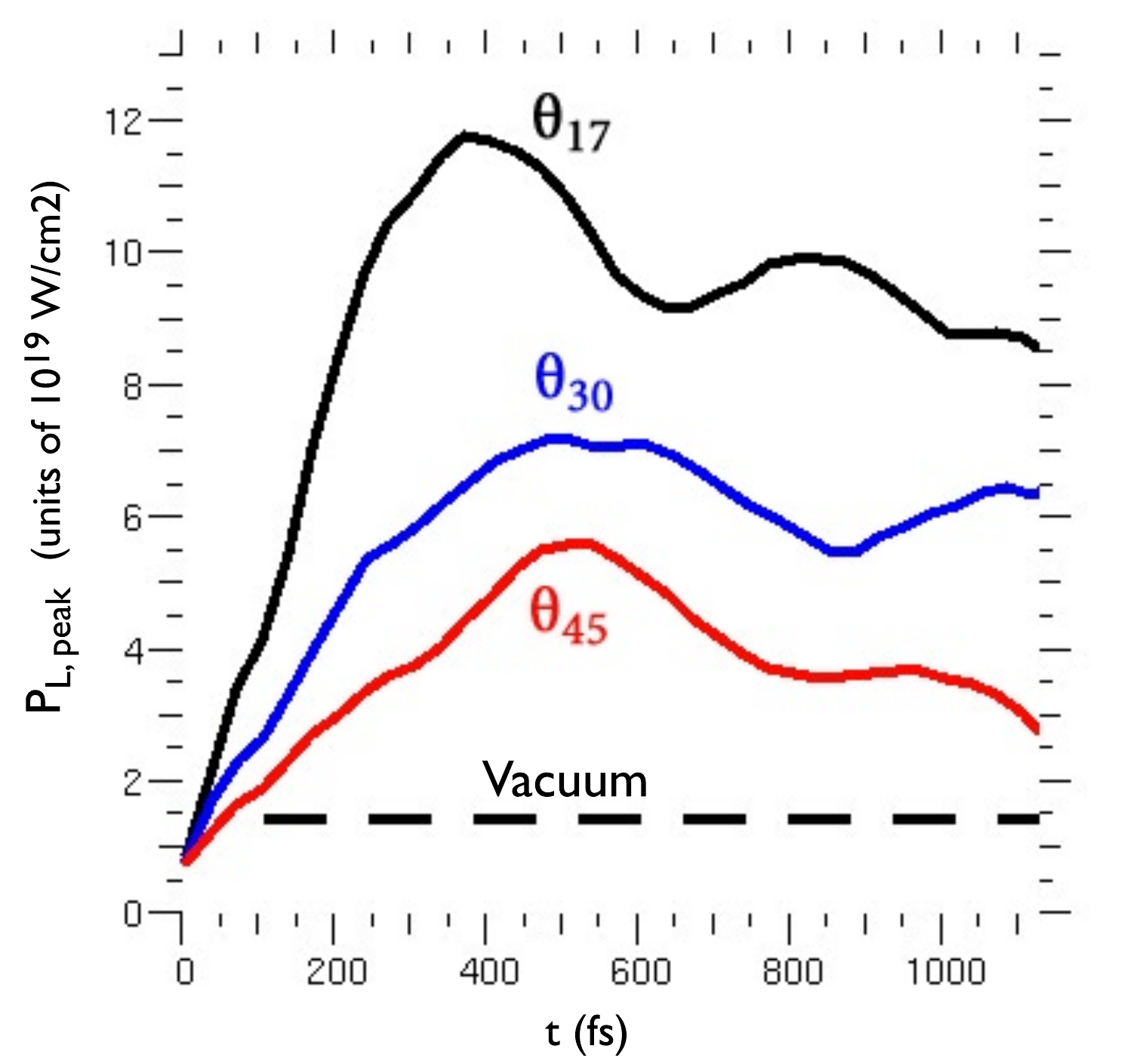}}
  \end{center}
  \caption{Laser peak intensity dependence on target angle for the \iII pulse (see text).}
  \label{fig:peakzt} 
\end{figure}

\begin{figure*}[t]
\begin{center}
\resizebox{16.5cm}{8.8cm}{\includegraphics{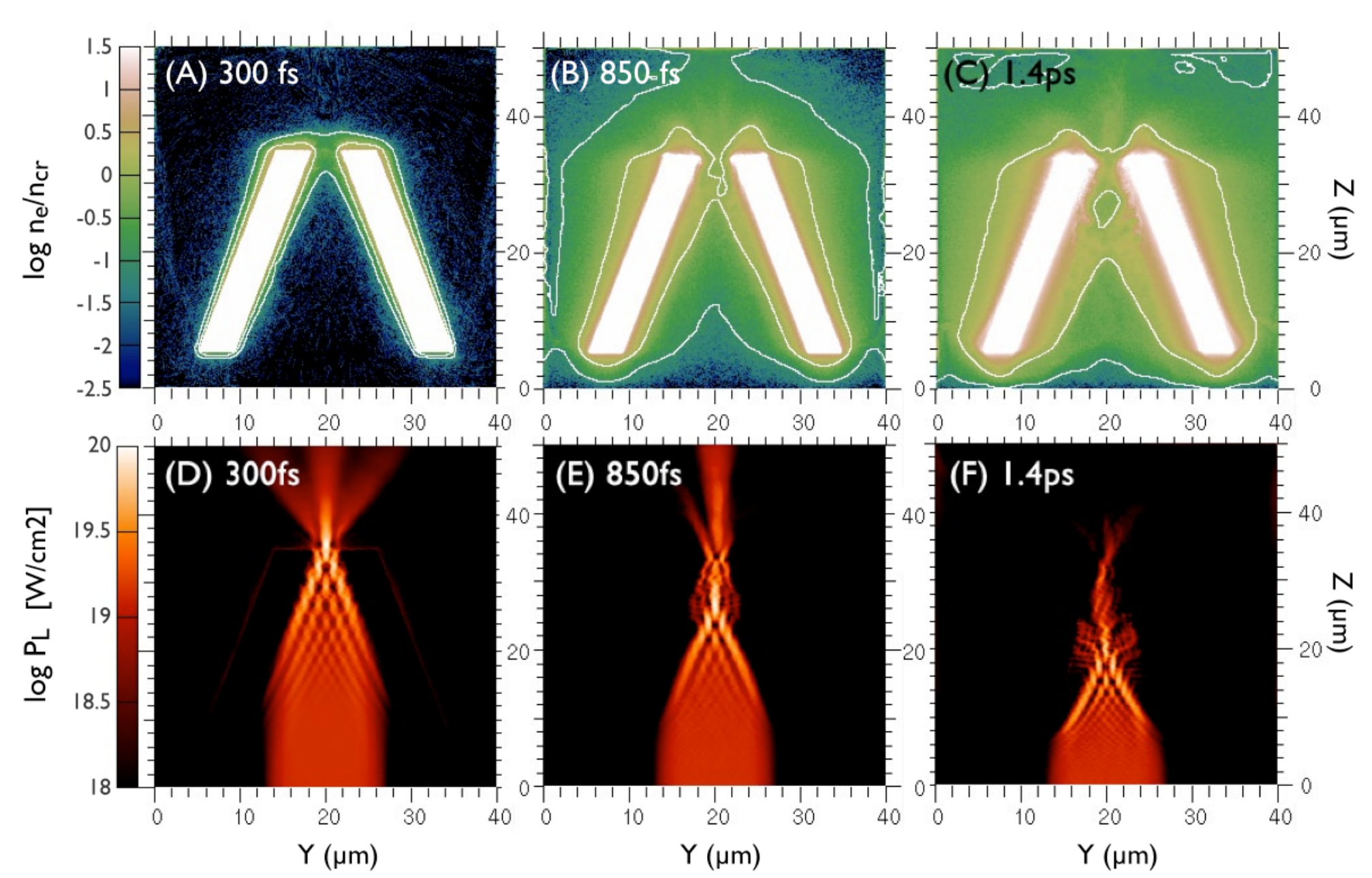}}
  \end{center}
  \caption{Temporal evolution of the \iII \tI pulse-target configuration in $Y-Z$ space. (A-C) Time-averaged electron density $\log n_e/n_{cr}$ with superimposed white contour lines corresponding to densities of \ecI and \ecII; (D-F)  time-averaged laser Poynting flux $\log P_L$ [$W/cm^2$].}
\label{fig:ampnet17}
\end{figure*}

\begin{figure*}[t]
\begin{center}
\resizebox{16.5cm}{5.5cm}{\includegraphics{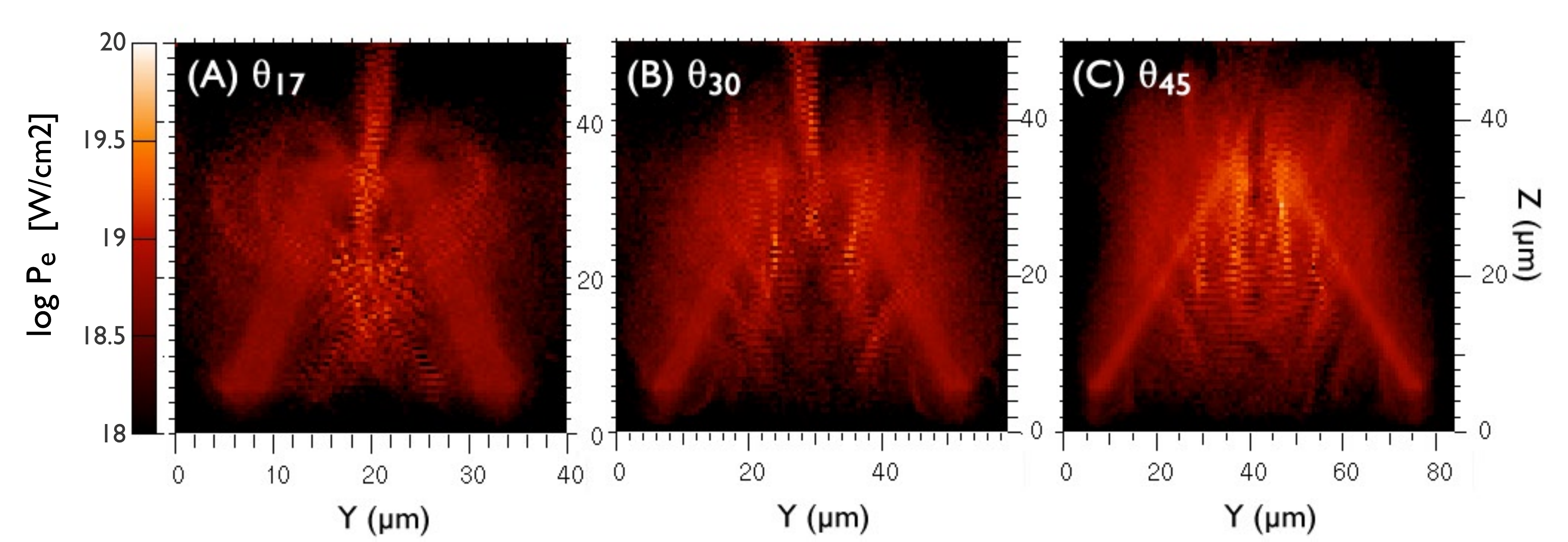}}
  \end{center}
  \caption{Hot electron characteristics at \cIII across target geometries for the \iII laser pulse; energy flux densities of forward-going electrons $\log P_e$ [$W/cm^2$] (note different transverse scales).}
\label{fig:hote}
\end{figure*}

Fig. \ref{fig:ampnet17} depicts the evolution of several key LPI processes for the \iII \tI pulse-target configuration.  As shown in Fig. \ref{fig:peakzt}, laser focusing is most efficient in this geometry where focusing at peak intensity $I_{\pt}/I_0\approx 8.8$ is achieved at time $t_{\pt} \approx$ 350 fs and sustained over 200 fs. Compared to an ideal optical case for \tI, i.e. an immobile ion target composition with $a\ll 1$ laser pulse as depicted in Fig. \ref{fig:setup}, this $I_{\pt}$ represents 82\% focusing efficiency.  Peak laser focusing in this optical simulation is in agreement with results from the comparable 18\dd \ target configuration reported on by Zeng \textit{et al.} to within 13\%.\cite{Zeng2010} The inclusion of additional effects from self-consistent evolution of the target walls shows that laser focusing due to geometric optical compression is both attenuated by collisionless laser energy coupling into particles and boosted by nonlinear LPI processes. As the propagation channel fills with ablated underdense plasma, laser focusing is enhanced as the pulse rapidly enters the regime of relativistic self-induced transparency regime for $P_L \ge P_c = 17 (\omega_0 / \omega_{pe})^2 $ GW\cite{Esarey2009}.  At early time Fig. \ref{fig:ampnet17} (A) shows surface plasma with density exceeding \ecI in the tip region ablated by collisionless laser absorption mechanisms.  Brunel (vacuum) heating\cite{Brunel1987} is dominant here due to the oblique angle of incidence; its signature is further evident in Fig. \ref{fig:hote} (A) at \cIII where hot electrons are bunched and driven through the target along the direction normal to its surface, with maximal energy flux density \per$\approx 3.2$.  In the higher intensity \iIII \tI pulse-target configuration, we observe identical values of \per$\approx 3.2$ and $I_{\pt}/I_0\approx 8.8$. 

\begin{figure}[t]
\begin{center}
\resizebox{8.2cm}{!}{\includegraphics{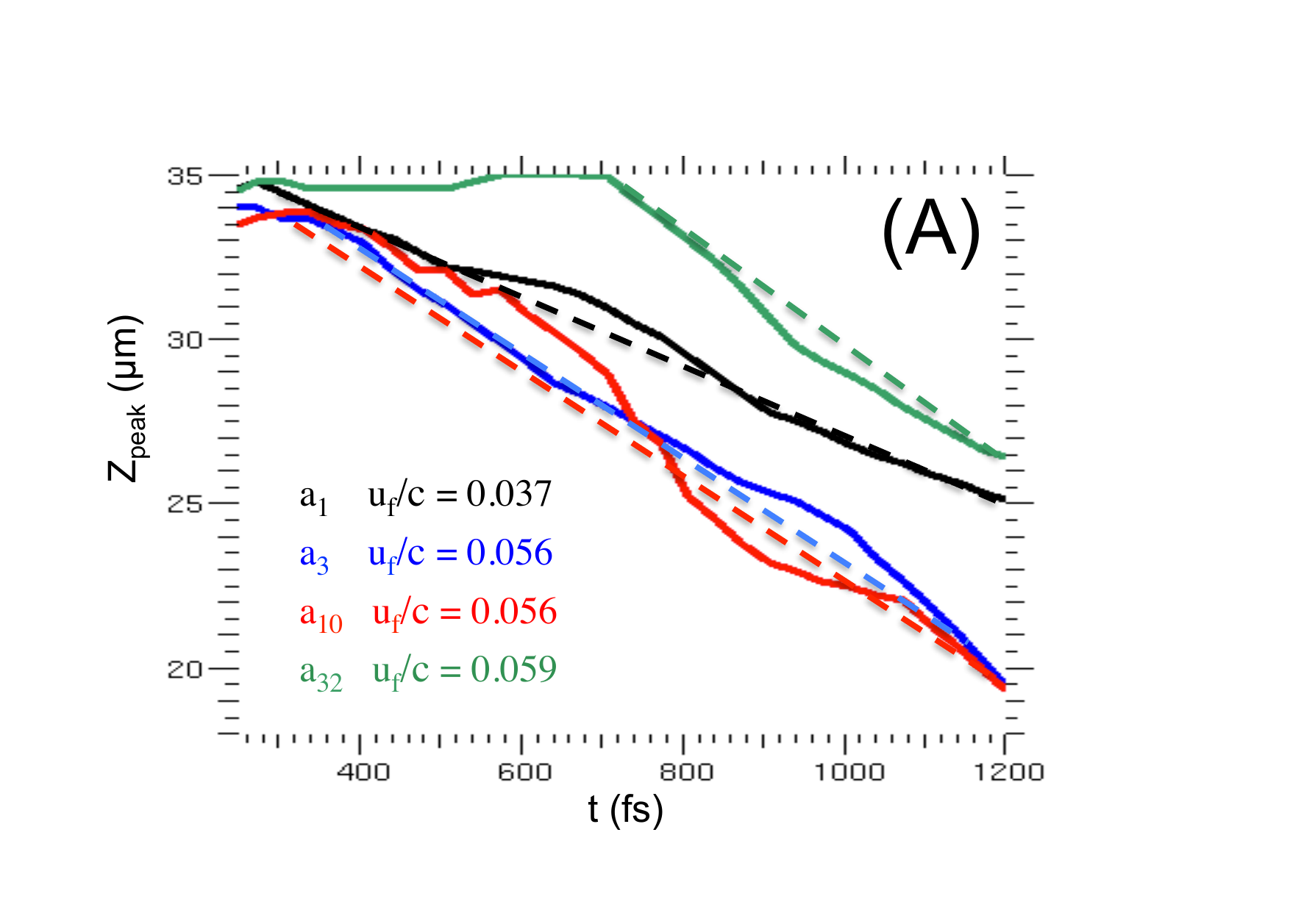}}
\resizebox{8.2cm}{!}{\includegraphics{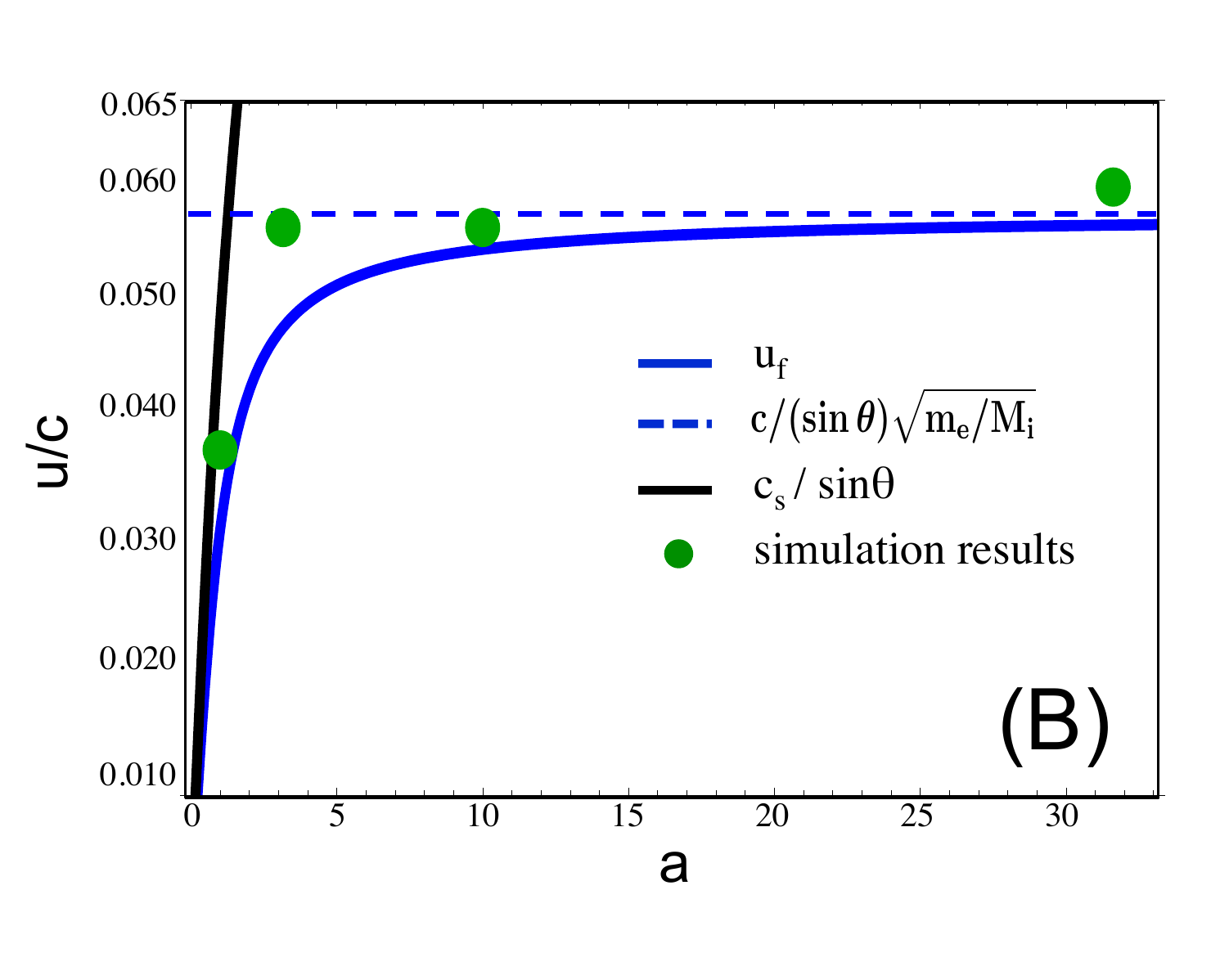}}
  \end{center}
  \caption{(A) Spatial location of peak laser intensity for the \tI target; front velocity $u_f$ and time evolution of $z_{peak}$ (see text). (B) Scaling of $u_f$ with dimensionless laser parameter $a$ in \tI. The analytic model from equation (\ref{eqn:ufangle}) and its asymptotic form are represented by the solid and dashed blue curves respectively.}
\label{fig:recess}
\end{figure}

We identify a clear trend in the \tI geometry where the region of peak laser intensity regresses away from the target tip as depicted in Fig. \ref{fig:recess}. While the magnitude of the the peak intensity $I_{\pt}/I_0$ decays over hundreds of femtoseconds, the effective laser focal point recedes from the tip with intensity-dependent rates $u_f \equiv - d z_{\pt}/dt\approx 0.037 c$ for \iI and $u_f \approx 0.056 c \ $ for both pulses \iII and \iIII.  Together with the constant \per and $I_{\pt}/I_0$ ratios observed for $a>1$ laser pulses, this trend suggests the \tI geometry may be representative of a broader class of configurations that are characterized by a sustained dynamical equilibrium between plasma expansion and laser radiation pressure.

\mcl{Here we present a simple but general hydrodynamical model for pulse-target configurations in which a recession velocity with $\frac{d}{dt} u_f=0$ is observed.  Consider an ion front moving at $u_f$ at the laser interface in planar geometry. The relevant conservation equations are given by,
\begin{eqnarray}
\partial_t \rho + \partial_z (\rho u) = 0 \label{eqn:m1} \\
\partial_t (\rho u) + \partial_z (\rho u^2 + P) = 0 \label{eqn:m2}
\end{eqnarray}
where $\rho$ is the mass density.  We now transform to a frame comoving with the ion front at $u_f$.  In this frame, the front is stationary and two oppositely-directed flows of equal magnitude are present.   By conservation of mass in equation (\ref{eqn:m1}) we see that the densities associated with these flows must be equal.  Electrons in the critical interface are heated by the laser to $T_e \simeq T_{pond.}$ and particles are effectively trapped by $p_z<0$ plasma flowing into the laser channel.  This quasi-isothermal interface contributes a plasma pressure term to the equation of motion and by equation (\ref{eqn:m2}) we may write,
\begin{equation}
u_f^2 M_i n_i = I/c - n_i T_e \label{eqn:upb}
\end{equation}
or,
\begin{equation}
u_f = \big( u_{w}^2  - c_s^2\big)^{1/2} \label{eqn:uf0}
\end{equation}
where
\begin{eqnarray}
c_s = (T_{h}/M_i)^{1/2} \label{eqn:cs} \\
\frac{u_{w}}{c} = \bigg(\frac{I}{M_i n_i c^3} \bigg)^{1/2} = \bigg(\frac{Z m_e a^2}{2 \gamma M_i} \bigg)^{1/2}
\label{eqn:uw}
\end{eqnarray}
Here equation (\ref{eqn:upb}) assumes full reflection and $T_e\gg T_i$ in the interface. Equation (\ref{eqn:cs}) is the ion-acoustic velocity with the ponderomotive temperature scaling, and equation (\ref{eqn:uw}) represents the ion front velocity from \wal \cite{Wilks1992}}

\mcl{Evaluating equation (\ref{eqn:uf0}) using (\ref{eqn:cs}-\ref{eqn:uw}) yields,
\begin{equation}
\frac{u_f}{c} = \bigg[\frac{m_e}{M_i} \big( 1 - \gamma^{-1} \big)\bigg]^{1/2}  \label{eqn:uf1}
\end{equation}
}

\mcl{This expression represents the front velocity for planar geometry at normal incidence. In our converging target geometry, the front will recede towards the injection plane at a higher rate as the plasma flows over a shorter distance into the laser channel. Unbalanced by laser pressure, a straight-forward calculation shows the recession velocity for a converging target with angle $\theta$, $u_f=c_s/\sin \theta$ reaches $0.17c$ for \iIII.  Transforming to the appropriate frame and solving for $u_f$ yields, 
\begin{equation}
\frac{u_f}{c}  = \frac{1}{ \sin \theta} \bigg[\frac{m_e}{M_i} \big( 1 - \gamma^{-1} \big)\bigg]^{1/2}
\label{eqn:ufangle}
\end{equation}}

\mcl{As illustrated in Fig. \ref{fig:recess} (B), equation (\ref{eqn:ufangle}) accurately predicts the front velocity and demonstrates the asymptotic form of $u_f$ for $a>3$ laser pulses. Results from an additional simulation performed at higher laser intensity  ($I=1.37\times 10^{21} \ W/cm^2$) are consistent with the asymptotic value of 
\begin{equation}
\frac{u_f}{c}  = \frac{1}{ \sin \theta}\bigg(\frac{m_e}{M_i}\bigg)^{1/2} 
\label{eqn:ufasymp}
\end{equation}
 as depicted in in Fig. \ref{fig:recess}.} That the laser front feels zero net pressure over time implies a scaling $I_{\pt}, P_e \propto a^2$ due to the strong correlation between the two variables for $a>1$.  The nature of the coefficients here is determined by geometric, laser absorption and nonlinear focusing effects, and shall be the subject of examination in future work. In the context of $K_\alpha$ radiography this unique class of wedge targets could be employed to achieve tightly-focused spots with order-of-magnitude higher intensity than the incident beam, scaling to intensities beyond $a>10$ without degradation in performance.

In the $a\le 1$ regime, it is expected that laser focusing will be enhanced due to sharply reduced absorption, e.g. by 28\% for vacuum heating for \iI versus \iII using Brunel's 1-D analytical model\cite{Brunel1987}, while the critical power threshold for self-focusing in the underdense region is still readily achieved.  This is in good agreement with the results obtained for the \iI pulses, where we see higher peak focusing by a factor of 1.5 on average across geometries. In this regime the laser is not sufficiently relativistic to excite electrons up to MeV energies; at intensity \iI, $T_h=0.2$ MeV and we accordingly observe that $P_e=0$ across geometries. By contrast, electron energy flux densities in the center of Fig. \ref{fig:hote} (A) represent the convolution of several heating mechanisms that operate efficiently in the underdense plasma upstream of the target tip. SM LWFA, a process where the laser self-interaction leads to longitudinal modulations and electron bunching at $\omega_{pe}$, plays the dominant role in particle heating here. For the inhomogenous underdense plasma present in Fig. \ref{fig:ampnet17} (B-C), the longitudinal electric field generated in this mechanism produces a broad continuous electron temperature spectra ranging from keV to hundreds of MeV\cite{Esarey2009}.  The inverse free electron process also contributes to particle heating; in this mechanism quasistatic electric and magnetic fields in the laser channel induce transverse betatron oscillations that couple laser energy into electrons when the betatron frequency $\omega_\beta \approx \omega_{pe}/2\gamma^{1/2}$ nears resonance with the optical frequency in the frame of the relativistic electron\cite{Pukhov2003a}. The signature of this process is present in Fig. \ref{fig:hote} (A) where we observe longitudinal bunching of electrons at $2\omega_{pe}$ with temperature scaling approximately three times higher than ponderomotive.  In the \iII \tI and \iII \tII pulse-target configurations, electron dynamics at late times are characterized by the presence of a dominant hot electron filament aligned with the target tip. For these configurations \per$\approx 3.2,\ 1$ respectively, as depicted in Fig. \ref{fig:hote} (A-B) with scalings to higher intensity in \tI described above. The potential viability of these energetic collimated electron structures for particle beam systems will be discussed briefly in the next section.

\begin{figure}[t]
\resizebox{8.2cm}{!}{\includegraphics{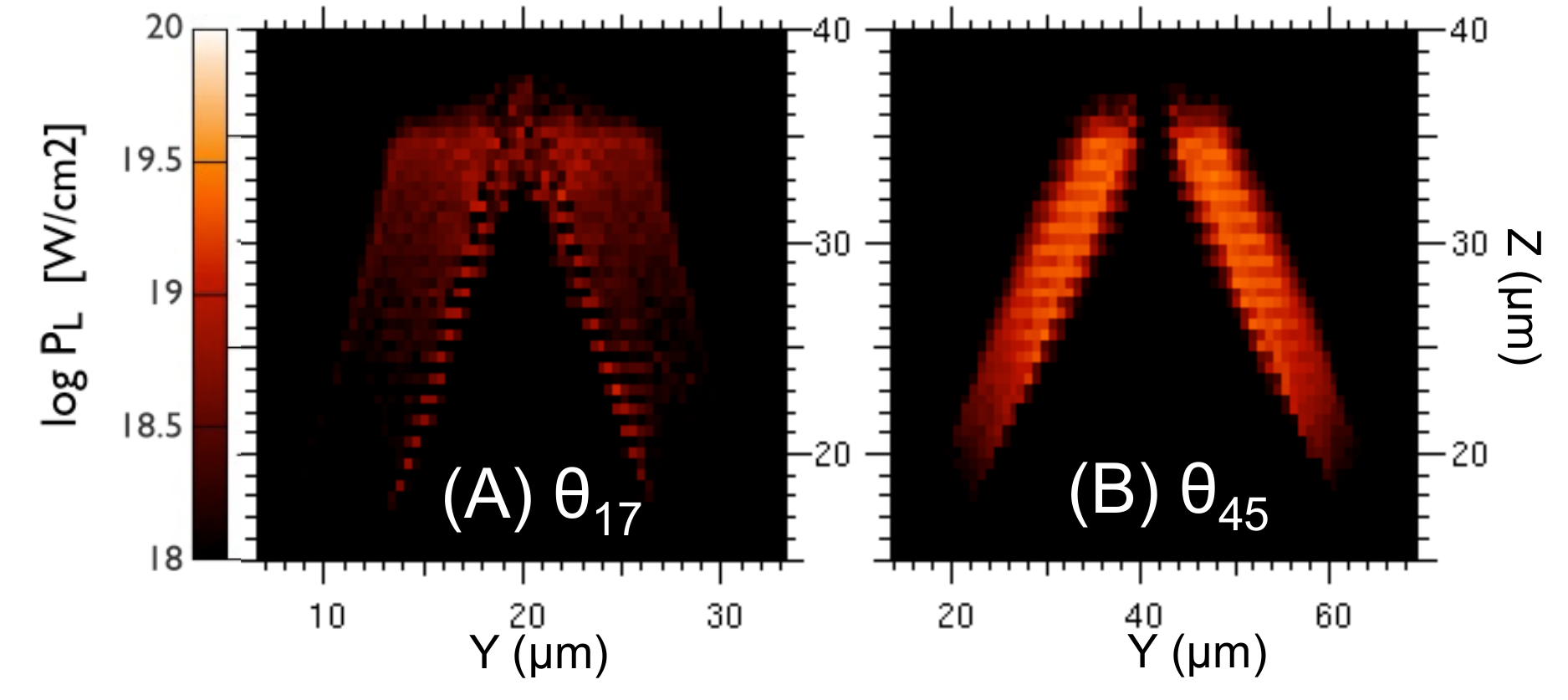}}
  \caption{Laser energy absorption into target walls at \cI for the \iII laser pulse (note different transverse scales).}
\label{fig:hoteearly}
\end{figure}

For the \iII \tII configuration, plasma ablation in the critical interface is comparable to \iII \tI, with focusing efficiency attenuated to $I_{\pt}/I_0\approx 5.2$ as shown in Fig. \ref{fig:peakzt}.  This peak is shifted to later time by 100 fs compared to \iII \tI due to the longer light propagation time and interaction region associated with the wider target angle. The \iII \tIII configuration at late times is characterized by \ecII plasma in the tip region and significant laser filamentation. This filamentation, coupled with the large spatial width of the target, suppresses ponderomotive self-channeling and results in effective closure of the target tip. The diminished peak focusing $I_{\pt}/I_0\approx 3.7$ is again shifted out to later time by $\approx 150$ fs as seen in Fig. \ref{fig:peakzt} by which time the tip has already started to close.  Fig. \ref{fig:hoteearly} illustrates the 11 times higher laser energy coupling into the target walls at \cI in this configuration compared to \iII \tI. This absorption has the effect of dephasing the laser light and, together with the lower geometrical compression associated with wider targets, plays an important role in limiting the attainable peak laser focusing in the \tII and \tIII geometries.  For the \iIII \tII configuration, peak focusing of $I_{\pt}/I_0\approx 6.9$ is sustained over a 600 fs interval prior to tip closure at $t\approx 1.1$ ps due to the attenuated ponderomotive force associated with laser filamentation. In addition we find electron acceleration in the tip region in this configuration to be effectively suppressed. As depicted in Fig. \ref{fig:hote} (C) the \iII \tIII pulse-target configuration exhibits two equally dominant hot electron filaments off to the sides of the tip with negligible electron acceleration through the tip. These structures are spatially coincident with strong filaments in the laser Poynting flux. This description is also representative of hot electron generation for \iIII \tIII. In these configurations, laser filamentation coupled with the large transverse width of the target suppresses laser pressure and the tip effectively closes.

\section{Conclusion}

We have employed two dimensional particle-in-cell simulations to study the focusing and evolution of ultraintense short pulse lasers in idealized converging targets. The wide 45\dd \ targets are found to exhibit generally suboptimal performance over time; for the $I_{19}$ and $I_{20}$ pulses, particle heating is characterized by two equally dominant hot electron filaments off to the sides of the tip. Ponderomotive self-channeling is suppressed by laser filamentation and the target tip effectively closes at $t\approx 450$ fs.  In the 30\dd \ geometry performance degrades with increasing intensity; for the $I_{19}$ pulse, peak laser focusing of $I_{\pt}/I_0\approx 5.2$ and the generation of tip-aligned electron filaments with \per$\approx 1$ are observed. In the $I_{20}\ - $ 30\dd \ configuration, sustained peak focusing of $I_{\pt}/I_0\approx 6.9$ is present at early times, with electron acceleration in the tip region suppressed as the tip closes at $t\approx 1.1$ ps. In general, performance in these wider targets is limited by large interaction regions and high absorption in the overdense walls that act to dephase laser light, diminished geometrical compression with respect to the 17 \dd \ target, and instabilities such as laser filamentation that deleteriously affect ponderomotive self-channeling.

By contrast, in the 17\dd \ configuration both peak laser focusing of $I_{\pt}/I_0\approx 8.8$ and electron generation and propagation through the target tip with \per$\approx 3.2$ are achieved for $a>1$. This peak focusing represents 82\% efficiency relative to an ideal optical case, comparable to results reported on previously\cite{Zeng2010}, with losses primarily going into hot electron excitation. A clear trend is described for this geometry where the region of peak laser intensity regresses away from the target tip at intensity-dependent rates that saturate for pulses of $a>3$. This trend suggests the 17\dd \ geometry may be representative of a broader class of pulse-target configurations that are characterized by a sustained dynamical equilibrium between  plasma expansion and laser radiation pressure. The physical picture here centers on the interface between the receding laser front and the plasma expansion.  The onset of recession occurs with a density perturbation in this region at early time, and as the plasma expands ablated particles stream in to the laser channel. Over time the laser energy density gradient becomes gentler, outward ponderomotive pressure decreases, and the laser front regresses towards injection plane. A simple but general model is developed that encapsulates the essential character of these systems which accurately predicts $u_f$ and yields key scalings $I_{\pt}, P_e \propto a^2$. For the parameters considered here, the hydrodynamic plasma expansion breaks down beyond subpicosecond timescales after which the high collisionality, ionization and resistive effects that dominate deep in realistic high-Z composition target walls must be included. Future research will focus on \mcl{expanding the applications of our model to a broader set of configurations}, and on analytically identifying pulse-target configurations in this dynamical equilibrium between plasma expansion and ponderomotive cavitation. Our results suggest the 17\dd \ and other configurations of this class may be employed as an experimentally straight-forward apparatus to achieve bursts of 10x boosted light for hundreds of femtoseconds, focused down to small spots of $\sim 1 \lambda_0$ with high performance scaling beyond intensities $a>10$.  

In addition the results presented here may have relevance for production cone target environments such as fast ignition. Although our simulation geometry is Cartesian and therefore ``slab,'' this is a reasonable facsimile of the physics that will occur for conical targets in $R-Z$ space. Assuming a linear scaling for geometrical compression and given laser \textit{p}-polarization, focusing for these configurations in three dimensions is enhanced by an additional factor of 4. In the 17\dd \ geometry this corresponds to a $\sim$35 times higher intensity spot in a conical target with a collimated, tip-aligned hot electron filament for $a>1$ laser pulses over $\sim 500$ fs.  Though electron transport in the dense thermonuclear fuel remains a key outstanding issue in the FI literature, the enhanced laser focusing described in this paper could in principle allow ignitor pulse requirements to met near $I_{19}$. Furthermore, for a given amount of input laser energy, these configurations could enable an additional control parameter to tune beam intensity at the tip of target. If higher intensity pulses are required, the saturation of $u_f$ potentially enables current cone-wire experiments\cite{Nakamura2009} to utilize narrow targets for gains in laser focusing while suffering no additional penalty in laser-fuel payload distance. 

This work was performed under the auspices of the U.S. Department of Energy by Lawrence Livermore National Laboratory under Contract DE-AC52-07NA27344. \mcl{The authors thank the anonymous referee for comments helpful to the polishing of the manuscript.} The first author would like to acknowledge funding and guidance provided by LLNL's Institute for Laser Science Applications (https://ilsa.llnl.gov). Computational effort for this work was supported by the LLNL Institutional Computing Grand Challenge program.

\newpage


\newpage

\end{document}